\documentclass[nofootinbib,a4paper,reprint,12pt,aps,tightenlines,onecolumn,longbibliography]{revtex4-1}
\usepackage[utf8]{inputenc}
\usepackage{amsmath}
\usepackage{amssymb}
\usepackage{cancel}
\usepackage{stmaryrd}
\usepackage{amsfonts}
\usepackage{bm}
\usepackage{bbm}
\usepackage{color}
\usepackage{graphicx}
\DeclareSymbolFont{operators}{OT1}{cmr}{m}{n}
\DeclareSymbolFont{letters}{OML}{cmm}{m}{it}
\DeclareSymbolFont{symbols}{OMS}{cmsy}{m}{n}
\DeclareSymbolFont{largesymbols}{OMX}{cmex}{m}{n}
\usepackage[hyperindex,breaklinks]{hyperref}
\hypersetup{
    pdfborder={0 0 0},
    allbordercolors={0 0 0},         
    unicode=false,          
    pdftoolbar=true,        
    pdfmenubar=true,        
    pdffitwindow=false,     
    pdfstartview={FitH},    
    pdfcreator={Eero Hirvijoki},
    pdfproducer={pdflatex},
    pdfnewwindow=true,
    colorlinks=true,
    linkcolor=blue,
    citecolor=blue,
    filecolor=blue,
    urlcolor=blue,  
}
\usepackage{array}
\usepackage{amsthm}
\usepackage{amsmath}

\newcommand{\grad}{\ensuremath{\text{grad}}}
\newcommand{\curl}{\ensuremath{\text{curl}}}
\renewcommand{\div}{\ensuremath{\text{div}}}

\makeatother

\begin{document}

\title{Charge-conserving, variational particle-in-cell method for the drift-kinetic Vlasov-Maxwell system}

\author{Eero Hirvijoki}
\affiliation{Department of Applied Physics, Aalto University, P.O. Box 11100, 00076 AALTO, Finland}
\email{eero.hirvijoki@gmail.com}

\date{\today}

\begin{abstract}
    This paper proposes a charge-conserving, variational, spatio-temporal discretization for the drift-kinetic Vlasov-Maxwell system, utilizing finite-elements for the electromagnetic fields and the particle-in-cell approach for the Vlasov distribution. The proposed scheme is fully electromagnetic, dealing with fields instead of potentials, and includes the effects of polarization and magnetization in the Gauss and Amp\`ere-Maxwell laws, a consequence of reducing the full particle dynamics to drift-center dynamics. There is, however, no need to invert the Gauss law: it is satisfied automatically at every time-step as a result of a discrete Noether symmetry, and the electric field is updated directly from the Amp\`ere-Maxwell equation. The method provides an update for the magnetic field that is fully explicit, involving only local operations. The update for particles is implicit for each particle individually, also leading to local operations only. The update for the electric field is linearly implicit due to the presence of a finite-element mass matrix and polarization and magnetization effects in the Amp\`ere-Maxwell equation, hence involving a sparse matrix inversion once at every time step. Because the scheme deals with the electromagnetic fields and not the potentials, it also provides the first serious attempt at constructing a structure-preserving numerical scheme for the mixed kinetic-ion--drift-kinetic-electron Vlasov-Maxwell model. Consequently, the proposed method could be used to simulate electromagnetic turbulence in fusion experiments or space plasmas that exhibit a strong background magnetic field while retaining all of the ion physics, most of the necessary electron physics, yet eliminating perhaps the biggest obstacle in reaching macroscopic transport time scales in kinetic simulations, namely the electron cyclotron time scale.
\end{abstract}

\maketitle 

\section{Introduction}
In the recent years, massive leaps have been taken in understanding and developing structure-preserving algorithms for simulation of plasmas (see \cite{Morrison-review:2017PhPl} for a recent review and the exhaustive list of references therein). At the forefront of this development have been the so-called geometric particle-in-cell (GEMPIC) methods \cite{Squire-Qin-Tang-PIC:2012PhPl,Evstatiev-shadwick:2013JCoPh,Shadwick-Stamm-Evstatiev:2014PhPl,Stamm-Shadwick-Evstatiev:2014ITPS,Xiao-et-al-kinetic:2015PhPl,He-et-al-Hamiltonian-splitting:2015PhPl,Qin-et-al:2016NucFu,Xiao-et-al-fluid:2016PhPl,Kraus-et-al:2017JPlPh,Xiao-et-al:2018PlST} that have had a profound impact on numerical simulation of both kinetic and two-fluid models. Based on discretizing either the underlying variational or Hamiltonian structure, GEMPIC algorithms provide unrivalled long-time fidelity and stability. This is especially important for kinetic simulations of fusion experiments where reaching macroscopic transport time scales of the order of $10^{-6} \text{s}$ requires a breathtaking number of time steps to resolve the electron cyclotron motion typically appearing at the order of  $10^{-11} \text{s}$. 

On par in terms of computational cost and complexity with other sophisticated energy- and/or charge-conserving particle-in-cell schemes based on direct discretization of the equations of motion \cite{Chen:2011jf,Chacon:2013cz,Chen:2014eh,Chacon:2016gi,Markidis:2011ep,Lapenta:2017uy,Chen-et-al:2019arXiv190301565C}, the GEMPIC methods preserve also the multisymplectic structure which typically has as many degrees-of-freedom as there are particles in the simulation. Discretizing the variational structure instead of the equations of motion is advantageous also from the point of investigating local energy-momentum conservation laws which in the infinite-dimensional time-continuous limit result from the translational and rotational Noether symmetries of the action functional. A fully discrete analog of such local symmetries has been realized for the pure Maxwell action in vacuum \cite{Xiao-et-al-lattice-Maxwell:2019PhLA} and the existence of a spatially discrete, local energy conservation law for the Vlasov-Maxwell GEMPIC encourages future studies into the topic \cite{Xiao-el-al-local-energy:2017PhPl}. 

In the midst of a florry of new work dealing with the full-particle Vlasov-Maxwell system, little attention has been paid to the reduced Vlasov-Maxwell plasma models, namely the guiding-center \cite{Brizard-Tronci:2016,Sugama_et_al_2018PhPl} and the gyrokinetic Vlasov-Maxwell models \cite{Burby-et-al:2015PhLA,Burby-Brizard:2019PhLA}. Although the rather peculiar choice of local velocity-space coordinates for particle motion in these models leads to more involved expressions than what is encountered in the full-particle Vlasov-Maxwell system -- non-linear polarization and magnetization may appear in both the Amp\`ere-Maxwell and the Gauss' laws -- the payoff is the inherent elimination of perhaps the biggest obstacle on the way to macroscopic transport time scales in fusion-experiment simulations, namely the cyclotron motion. Several studies have focused on variational integration of individual guiding-center motion \cite{Qin-Guan:2008PhRvL,Squire-et-al-guiding-center:2012PhPl,Kraus-projected-degenerate:2017arXiv,Ellison-Burby:2017PhPl,Ellison-et-al-degenerate:2018PhPl} but a structure-preserving discretization of the full system, including the evolution of the electric and magnetic fields, has not been presented as of yet. The few existing self-consistent studies have either stopped at deriving a finite-dimensional noncanonical Poisson bracket and a Hamiltonian for the reduced models \cite{Burby-finite-dimensional-2017PhPl}, or used standard numerical methods for first order ordinary differential equations to advance the finite-dimensional Hamiltonian system forward in time \cite{Evstatiev-guiding-center-vlasov-maxwell:2014CoPhC}, a step which does not guarantee the preservation of the underlying mathematical structure. Specifically, it has been anticipated that the complications of non-locality in gyrokinetics and the nonlinear effects of polarization and magnetization in the non-canonical Poisson bracket of the finite-dimensional system might make it difficult to find symplectic integrators based on traditional Hamiltonian splitting techniques.

Since the reward of eliminating the electron cyclotron motion from kinetic simulations of magnetized plasmas is a speed-up of at least three orders of magnitude, we will not give up on this task without a fight. Instead, we target the drift-kinetic Vlasov-Maxwell model that is obtained as the long-wave-length limit of the gyrokinetic Vlasov-Maxwell model, and derive a structure-preserving, variational, spatio-temporal algorithm for the drift-kinetic system with an exact charge conservation law. The choice to study the drift-kinetic model is a conscious one and aims at avoiding the non-localities present in gyrokinetics. Furthermore, since existing gyrokinetic particle-in-cell codes in practice already use a time-step that is close to the ion cyclotron period, and simulations in the steep gradients in the plasma edge and scrape-off layer might in fact require kinetic treatment of ions, we see that a structure-preserving discretization of the drift-kinetic Vlasov-Maxwell system that can be coupled to structure-preserving treatment of fully kinetic ions would be in order.

The foundation for our new algorithm lies in the recently developed gauge-free gyrokinetic theory \cite{Burby-Brizard:2019PhLA}. Without it, it would likely be impossible to formulate the model in terms of the electric and magnetic fields only, and to couple the resulting drift-kinetic system with fully kinetic ions. This is reflected in the fact that variational electromagnetic gyrokinetic theories typically involve the potentials in the equations of motion for a single gyrocenter. It is thus natural to begin the current exposition with a recap of the time- and space-continuous, electromagnetically gauge-invariant drift-kinetic Vlasov-Maxwell system. After the review, we discuss the discretization of the model in two steps. The spatial discretization turns the infinite-dimensional system into a finite-dimensional one which is electromagnetically gauge invariant similarly as the infinite-dimensional system. The temporal discretization then provides an update map for the magnetic field that is fully explicit, an update map for the drift-center coordinates that is implicit for each drift-center individually, and an update map for the electric field that is linearly implicit due to the presence of a finite-element mass matrix and polarization and magnetization in the Amp\`ere-Maxwell equation. Finally, we demonstrate how a discrete Noether symmetry related to the electromagnetic gauge invariance leads to exact charge conservation law and guarantees the preservation of Gauss' law during the temporal advance. 
\section{The drift-kinetic Vlasov-Maxwell system}
We begin with a recap of the drift-kinetic Vlasov-Maxwell model that is obtained as the long-wave-length limit of the gyrokinetic Vlasov-Maxwell system \cite{Burby-Brizard:2019PhLA}. As the combination "Vlasov-Maxwell" in the name suggests, the model is effectively an infinite-dimensional first-order system of ordinary differential equations for the fields $(F,\bm{E},\bm{B})$. These fields are the familiar electric and magnetic field, $\bm{E}(t,\bm{x})$ and $\bm{B}(t,\bm{x})$, and the drift-center phase-space density distribution $F(t,\bm{x},u;\mu)$ that depends locally on the drift-center phase-space coordinates, namely the spatial position $\bm{x}$, the parallel velocity $u$, and the magnetic moment $\mu$. The dependency on the magnetic moment is parametric, not dynamic, a detail that becomes clear soon. The spatial density, counting the number of drift-centers per cubic meter, is defined as $n(\bm{x})=\int F(\bm{x},u;\mu)du d\mu $. 

The first-order dynamical nature of the system is revealed once the evolution equations are presented
\begin{align}
    \partial_t F+\nabla\cdot(\bm{\xi}^{x}F)+\partial_u(\xi^uF)&=0,\\
    \partial_t\bm{D}-\nabla\times\bm{H}+\bm{j}&=0,\\
    \partial_t\bm{B}+\nabla\times\bm{E}&=0,\\
    \nabla\cdot\bm{D}-\varrho&=0,\\
    \nabla\cdot\bm{B}&=0.
\end{align}
The components of the drift-center phase-space velocity field $(\bm{\xi}^x,\xi^u)$ are given by
\begin{align}
    \bm{\xi}^x&=\frac{\bm{B}^{\star}}{m\bm{b}_{\text{ext}}\cdot\bm{B}^{\star}}\frac{\partial G}{\partial u}+\frac{(q\bm{E}-\nabla G)\times\bm{b}_{\text{ext}}}{e\bm{b}_{\text{ext}}\cdot\bm{B}^{\star}},\\
    \xi^u&=\frac{\bm{B}^{\star}\cdot(q\bm{E}-\nabla G)}{m\bm{b}_{\text{ext}}\cdot\bm{B}^{\star}},
\end{align}
with $\bm{B}^{\star}=\bm{B}+\bm{B}_{\text{ext}}+(m/q)u\nabla\times\bm{b}_{\text{ext}}$, and the macroscopic fields $\bm{D}$ and $\bm{B}$ and the "free" sources $\bm{j}$ and $\varrho$ are constructed from the trio $(F,\bm{E},\bm{B})$ according to
\begin{align}
    \bm{D}&=\varepsilon_0\bm{E}-\int \frac{\partial K}{\partial \bm{E}} F du d\mu,\\
    \bm{H}&=\mu_0^{-1}(\bm{B}_{\text{ext}}+\bm{B})+\int \frac{\partial K}{\partial \bm{B}}Fdud\mu,\\
    \bm{j}&=\int q\bm{\xi}^x F dud\mu,\\
    \varrho&=\int q F dud\mu.
\end{align}
The function $G(\bm{x},u,\bm{E},\bm{B};\mu)=E(\bm{x},u;\mu)+K(\bm{x},u,\bm{E},\bm{B};\mu)$, encoding much of the physics, is the sum of the guiding-center kinetic energy $E$ and the drift-kinetic perturbation term $K$, which are defined according to
\begin{align}
    E&=\frac{1}{2}m u^2+\mu|\bm{B}_{\text{ext}}|,\\
    K&=\mu \bm{b}_{\text{ext}}\cdot\bm{B}+ (\mu|\bm{B}_{\text{ext}}|-mu^2)\frac{\bm{B}\cdot\mathbf{1}_{\perp}\cdot\bm{B}}{2|\bm{B}_{\text{ext}}|^2}\nonumber
    \\
    &\quad-\frac{m\bm{E}\cdot\mathbf{1}_{\perp}\cdot\bm{E}}{2|\bm{B}_{\text{ext}}|^2}-\frac{mu\bm{E}\times\bm{b}_{\text{ext}}\cdot\bm{B}}{|\bm{B}_{\text{ext}}|^2}.
\end{align}
The dyad $\mathbf{1}_{\perp}=\mathbf{1}-\bm{b}_{\text{ext}}\bm{b}_{\text{ext}}$ denotes a projection in the direction perpendicular to the static external magnetic field $\bm{B}_{\text{ext}}(\bm{x})$ and $\bm{b}_{\text{ext}}=\bm{B}_{\text{ext}}/|\bm{B}_{\text{ext}}|$ is the corresponding unit vector.

The drift-kinetic system above has a rigorous mathematical foundation in the form of an action principle from which the dynamical equations and the constraints can be derived after applying the Euler-Poincar\'e reduction and the Hamilton's principle of least action. In the action, a single-drift-center phase-space Lagrangian is multiplied with a phase-space density of fixed-value drift-center labels, the product integrated over all of the drift-center phase space and a time interval, and then combined with the standard electromagnetic action to account for a self-consistent treatment of the electromagnetic potentials appearing in the single-drift-center Lagrangian. In such a system, the electromagnetic potentials and fields are treated as Eulerian variables and the role of the single-drift-center trajectory is to carry the fixed-value phase-space-density labels along the drift-center phase-space flow. The process effectively produces a modified version of the action integral that Low constructed for the full-particle Vlasov-Maxwell system \cite{Low:1958}.

In the mixed-variable formalism, the drift-kinetic Vlasov-Maxwell action is then given by 
\begin{align}\label{eq:Low-action}
    S[\bm{x}(t),u(t),\bm{A}(t),\phi(t);F_0]=\int_{t_1}^{t_2}L(\bm{A}(t),\dot{\bm{A}}(t),\phi(t),\bm{x}(t),\dot{\bm{x}}(t),u(t);F_0)dt.
\end{align}
with the corresponding Lagrangian provided by 
\begin{align}
    &L(\bm{A}(t),\dot{\bm{A}}(t),\phi(t),\bm{x}(t),\dot{\bm{x}}(t),u(t);F_0)
    \nonumber
    \\
    &=\int\frac{1}{2}(\varepsilon_0|\bm{E}(t,\bm{x})|^2-\mu_0^{-1}|\bm{B}_{\text{ext}}(\bm{x})+\bm{B}(t,\bm{x})|^2)d^3\bm{x}
    \nonumber
    \\
    &+\int\big[q(\bm{A}(t,\bm{x}(t))+\bm{A}_{\text{ext}}(\bm{x}(t)))\cdot\dot{\bm{x}}(t)-q\phi(t,\bm{x}(t))\big]F_0d^3\bm{x}_0du_0 d\mu
    \nonumber
    \\
    &+\int \big[m u(t)\bm{b}_{\text{ext}}(\bm{x}(t))\cdot\dot{\bm{x}}(t)-E(\bm{x}(t),u(t));\mu)\big]F_0d^3\bm{x}_0du_0d\mu
    \nonumber
    \\
    &-\int K(\bm{x}(t),u(t),\bm{E}(t,\bm{x}(t)),\bm{B}(t,\bm{x}(t));\mu)F_0d^3\bm{x}_0du_0d\mu.
\end{align}
The dynamic electric and magnetic field are defined via the standard relations $\bm{E}=-\dot{\bm{A}}-\nabla\phi$ and $\bm{B}=\nabla\times\bm{A}$, and it is assumed that the static external magnetic field has an associated vector potential $\bm{B}_{\text{ext}}=\nabla\times\bm{A}_{\text{ext}}$. In the expressions above, one is to interpret $\bm{x}(t)=\bm{x}(t;\bm{x}_0,u_0;\mu)$, $u(t)=u(t;\bm{x}_0,u_0;\mu)$, and $F_0=F_0(\bm{x}_0,u_0;\mu)$. Low's original action for the full-particle Vlasov-Maxwell system would be recovered by the replacements $G=\tfrac{1}{2}m|\bm{v}(t)|^2$, $u(t)\bm{b}_{\text{ext}}(\bm{x}(t))=\bm{v}(t)$, $du_0d\mu=d\bm{v}_0$, with the interpretations $\bm{x}(t)=\bm{x}(t;\bm{x}_0,\bm{v}_0)$, $\bm{v}(t)=\bm{v}(t;\bm{x}_0,\bm{v}_0)$, and $F_0=F_0(\bm{x}_0,\bm{v}_0)$.

The connection of this mixed-variable action to the drift-kinetic Vlasov-Maxwell system could be revealed by the Euler-Poincar\'e reduction. We do not discuss the details of that process here but summarize the main points. Where $\bm{x}(t;\bm{x}_0,u_0;\mu)$ and $u(t;\bm{x}_0,u_0;\mu)$ can be viewed as a map moving a single drift-center from the point $(\bm{x}_0,u_0)$ to a point $(\bm{x}(t),u(t))$ in time $t$, and $\dot{\bm{x}}(t;\bm{x}_0,u_0;\mu)$ and $\dot{u}(t;\bm{x}_0,u_0;\mu)$ as the corresponding Lagrangian time derivative of that map, the velocity field $(\bm{\xi}^x,\xi^u)$ in the Vlasov formulation is simply the Eulerian view of the Lagrangian time derivative of the drift-center trajectory. Consequently $F(t,\bm{x},u;\mu)$ is nothing but the fixed-value $F_0(\bm{x}_0,u_0;\mu)$ carried to the positions the drift-centers move in time $t$. This connection between the Lagrangian and Eulerian formulations is the basis for the particle-in-cell approach to solving the Vlasov equation: one samples the fixed density $F_0$ with a set of markers and then pushes the markers forward in time according to their flow, to carry the initial density forward in time.
\section{Spatial Discretization}
To obtain a finite-dimensional approximation of the drift-kinetic Lagrangian and action, let's assume we have some domain $\Omega\subset\mathbb{R}^3$ and a finite-dimensional discretization of the associated de Rham complex: we expect there to be the sets of basis functions $\{W^0_i\}_i$, $\{\bm{W}^1_j\}_j$, $\{\bm{W}^2_k\}_k$, and $\{W^3_\ell\}_\ell$ such that
\begin{align}
    \nabla W^0_i&=\grad_i^j\bm{W}^1_j,\\
    \nabla\times\bm{W}^1_j&=\curl_j^k\bm{W}^2_k,\\
    \nabla\cdot\bm{W}^2_k&=\div_k^{\ell}W^3_{\ell}.
\end{align}
We also assume there to be the associated matrices
\begin{align}
    %\int_\Omega W^0_{i_1}W^0_{i_2}d\bm{x}=M^0_{i_1i_2}\\
    \int_\Omega \bm{W}^1_{j_1}\cdot\bm{W}^1_{j_2}d\bm{x}=M^1_{j_1j_2},\\
    \int_\Omega \bm{W}^2_{k_1}\cdot\bm{W}^2_{k_2}d\bm{x}=M^2_{k_1k_2}.
\end{align}
Throughout the rest of the paper, we will adopt Einstein summation over the repeated superscript--subscript index pairs. Furthermore, the letters $i,j,k,\ell$ always refer to the corresponding finite-element spaces as denoted above. %The number of finite-element basis functions in the different bases is typically not equal, and the matrices $\grad,\curl,\div$ typically not square matrices. 

Because the basis functions satisfy the de Rham complex, we have that
\begin{align}
    0&=\nabla\times\nabla W^0_i=\grad_i^j\nabla\times\bm{W}^1_j=\grad_i^j\curl_j^k \bm{W}^2_k,\\
    0&=\nabla\cdot\nabla\times\bm{W}^1_j=\curl_j^k\nabla\cdot\bm{W}^2_k=\curl_j^k\div_k^\ell W^3_\ell,
\end{align}
which implies the following matrix identities
\begin{align}
    \curl_j^k\grad_i^j&=0,\\
    \div_k^\ell\curl_j^k&=0.
\end{align}
The spatial discretizations of the vector and scalar potential are then taken to be
\begin{align}
    \bm{A}_{\text{ext}}&=a_{\text{ext}}^j\bm{W}^1_j(\bm{x}),\\
    \bm{A}&=a^j(t)\bm{W}^1_j(\bm{x}),\\
    \phi&=\phi^i(t)W^0_i(\bm{x}),
\end{align}
implying the following expressions for the finite-dimensional electric and magnetic and fields
\begin{align}
    \bm{E}&=(-\dot{a}^j-\phi^i\grad_i^j)\bm{W}^1_j=e^j\bm{W}^1_j,\\
    \bm{B}&=a^j\curl_j^k\bm{W}^2_k=b^k\bm{W}^2_k,\\
    \bm{B}_{\text{ext}}&=a_{\text{ext}}^j\curl_j^k\bm{W}^2_k=b_{\text{ext}}^k\bm{W}^2_k.
\end{align}
Consequently, the discrete magnetic field will satisfy the identity $\partial_t\nabla\cdot\bm{B}=-\nabla\cdot\nabla\times\bm{E}=e^j\div_k^\ell\curl_j^k W^3_\ell=0$, meaning that if the degrees of freedom for $\bm{B}$ initially satisfy $b^k\div^{\ell}_k=0$, they will satisfy the condition for all times.
The fixed-value density distribution $F_0(\bm{x}_0,u_0;\mu)$ we sample with markers according to
\begin{align}
    F_0=\sum_p\delta(\bm{x}_0-\bm{x}_p(t_0))\delta(u_0-u_p(t_0))\delta(\mu-\mu_p),
\end{align}
where $(\bm{x}_p(t_0),u_p(t_0))$ are the initial phase-space coordinates for the drift-center marker trajectory $(\bm{x}_p(t),u_p(t))$. In practice, every marker should be weighted with a label $w_p$ accounting for the number of real particles the marker represents. Here we have, however, suppressed this factor for notational clarity. From here on, we will also use the tuples $\mathbbm{x}=\{\bm{x}_p\}_p$, $\dot{\mathbbm{x}}=\{\dot{\bm{x}}_p\}_p$, $\mathbbm{u}=\{u_p\}_p$, $\mathbbm{a}=\{a^j\}_j$, $\dot{\mathbbm{a}}=\{\dot{a}^j\}_j$ $\mathbbm{b}=\{a^k\}_k$, $\mathbbm{e}=\{e^j\}_j$, and $\phi=\{\phi^i\}_i$ to group together the degrees of freedom. Especially it is to be understood that $\phi$ now refers to the tuple of degrees of freedom, not the space-continuous electrostatic potential.  

Substituting the above expressions to the drift-kinetic Vlasov-Maxwell action functional, we obtain a new action functional
\begin{align}
    %&
    S[\mathbbm{x}(t),\mathbbm{u}(t),\mathbbm{a}(t),\phi(t)]%\nonumber
    %\\
    &=\int_{t_1}^{t_2}L(\mathbbm{x}(t),\dot{\mathbbm{x}}(t),\mathbbm{u}(t),\mathbbm{a}(t),\dot{\mathbbm{a}}(t),\phi(t))dt,
\end{align}
where the new Lagrangian is
\begin{align}
    &L(\mathbbm{x}(t),\dot{\mathbbm{x}}(t),\mathbbm{u}(t),\mathbbm{a}(t),\dot{\mathbbm{a}}(t),\phi(t))
    \nonumber
    \\
    &=\frac{\varepsilon_0}{2}(-\dot{a}^{j_1}-\phi^{i_1}\grad_{i_1}^{j_1})M^1_{j_1,j_2}(-\dot{a}^{j_2}-\phi^{i_2}\grad_{i_2}^{j_2})\nonumber
    \\
    &-\frac{\mu_0^{-1}}{2}(a^{j_1}+a_{\text{ext}}^{j_1})\curl_{j_1}^{k_1}M^2_{k_1k_2}\curl_{j_2}^{k_2}(a^{j_2}+a_{\text{ext}}^{j_2})
    \nonumber
    \\
    &+\sum_p \big[q (a^j+a_{\text{ext}}^j)\bm{W}^1_j(\bm{x}_p)\cdot\dot{\bm{x}}_p-q\phi^iW_i^0(\bm{x}_p)\big]
    \nonumber
    \\
    &+\sum_p\big[mu_p\bm{b}_{d,\text{ext}}(\bm{x}_p)\cdot\dot{\bm{x}}_p-E_d(\bm{x}_p,u_p;\mu_p\big]
    \nonumber
    \\
    &-\sum_pK_d(\bm{x}_p,u_p,\mathbbm{e},\mathbbm{b};\mu_p).
\end{align}
%$L=L_{\text{EM}}+L_{\text{INT}}+L_{\text{DK}}$ is again split into three parts. The expression for the electromagnetic Lagrangian becomes
%\begin{align}
%    L_{\text{EM}}(\dot{a},a,\phi)=&\frac{\varepsilon_0}{2}(-\dot{a}^{j_1}-\phi^{i_1}\grad_{i_1}^{j_1})M^1_{j_1,j_2}(-\dot{a}^{j_2}-\phi^{i_2}\grad_{i_2}^{j_2})\nonumber
%    \\
%    &-\frac{\mu_0^{-1}}{2}(a^{j_1}+a_{\text{ext}}^{j_1})\curl_{j_1}^{k_1}M^2_{k_1k_2}\curl_{j_2}^{k_2}(a^{j_2}+a_{\text{ext}}^{j_2})
%\end{align}
%the interaction Lagrangian becomes
%\begin{align}
%    L_{\text{INT}}(a,\phi,\bm{x},\dot{\bm{x}})&=\sum_p \big[e (a^j+a_{\text{ext}}^j)\bm{W}^1_j(\bm{x}_p)\cdot\dot{\bm{x}}_p-e_p\phi^iW_i^0(\bm{x}_p)\big]
%\end{align}
%and the drift-kinetic part becomes
%\begin{align}
%    L_{\text{DK}}(a,\dot{a},\phi,\bm{x},\dot{\bm{x}},u)=&\sum_p\big[mu_p(t)\bm{b}_{d,\text{ext}}(\bm{x}_p(t))\cdot\dot{\bm{x}}_p(t)\nonumber
%    \\&-(E_d(\bm{x}_p,u_p;\mu_p)+K_d(\bm{x}_p,u_p,e,b;\mu_p))\big]
%\end{align}
The expressions for $\bm{b}_{d,\text{ext}}$, $E_d$, and $K_d$ are given by
\begin{align}
    \bm{b}_{d,\text{ext}}&=\frac{\bm{B}_{d,\text{ext}}}{|\bm{B}_{d,\text{ext}}|},
    \\
    E_d&=\frac{1}{2}m u^2+\mu|\bm{B}_{d,\text{ext}}|,
    \\
    K_d&=K^\mathbbm{b}_k(\bm{x},\mu)b^k+b^{k_1}K^{\mathbbm{bb}}_{k_1,k_2}(\bm{x},u,\mu)b^{k_2}\nonumber
    \\
    &\quad-e^{j_1}K^{\mathbbm{ee}}_{j_1,j_2}(\bm{x})e^{j_2}-e^jK^{\mathbbm{eb}}_{j,k}(\bm{x},u)b^k,
\end{align}
and we have introduced the functions
\begin{align}
    K^\mathbbm{b}_k&=\mu\bm{b}_{d,\text{ext}}\cdot\bm{W}^2_k,\\
    K^{\mathbbm{bb}}_{k_1,k_2}&=(\mu|\bm{B}_{d,\text{ext}}|-mu^2)\frac{\bm{W}^2_{k_1}\cdot\mathbf{1}_{d,\perp}\cdot\bm{W}^2_{k_2}}{2|\bm{B}_{d,\text{ext}}|^2},\\
    K^{\mathbbm{eb}}_{j,k}&=\frac{mu\bm{W}^1_j\times\bm{b}_{d,\text{ext}}\cdot\bm{W}^2_k}{|\bm{B}_{d,\text{ext}}|^2},\\
    K^{\mathbbm{ee}}_{j_1,j_2}&=\frac{m\bm{W}^1_{j_1}\cdot\mathbf{1}_{d,\perp}\cdot\bm{W}^1_{j_2}}{2|\bm{B}_{d,\text{ext}}|^2},
\end{align}
together with an expression for the discrete external magnetic field $\bm{B}_{d,\text{ext}}=b_{\text{ext}}^k\bm{W}^2_k$ and the associated projective dyad  $\mathbf{1}_{d,\perp}=\mathbf{1}-\bm{b}_{d,\text{ext}}\bm{b}_{d,\text{ext}}$. %This representation is chosen  it transparent that $K_d$ is at most quadratic in the variables $(\mathbbm{e},\mathbbm{b})$.

The finite-dimensional Lagrangian $L(\mathbbm{x},\dot{\mathbbm{x}},\mathbbm{u},\mathbbm{a},\dot{\mathbbm{a}},\phi)$ is electromagnetically gauge-invariant in the sense that, if we choose some $\chi=\chi^i(t)W^0_i(\bm{x})$ and make the changes
\begin{align}
    &a^j\rightarrow a^j+\chi^i\grad_i^j,\\
    &\phi^i\rightarrow \phi^i-\dot{\chi}^i,
\end{align}
the Lagrangian changes to 
\begin{align}
    L\rightarrow &L+\sum_p \big[q\dot{\bm{x}}_p\cdot \chi^i\grad_i^j\bm{W}^1_j(\bm{x}_p)+q\dot{\chi}^iW_i^0(\bm{x}_p)\big],\nonumber
    \\
    =&L+\frac{d}{dt}\Big[\sum_p q\chi^iW^0_i(\bm{x}_p)\Big].
\end{align}
Previously, this gauge freedom has been used in conjunction with the $\phi=0$ gauge to express the Lagrangian as a pure phase-space form and to derive the corresponding finite-dimensional Poisson bracket and Hamiltonian \cite{Burby-finite-dimensional-2017PhPl}. It was, however, anticipated that finding a Hamiltonian splitting scheme for advancing the system in time would perhaps be difficult to obtain due to the non-polynomial nature of the finite-dimensional Poisson-bracket with respect to the degrees-of-freedom. Hence, instead of repeating the analysis of the Hamiltonian structure, we consider temporal discretization of the action directly. %We remark that the system should retain a polynomial dependency on the degrees-of-freedom, if one assumes the external magnetic field to be uniform. 

\section{Temporal discretization and Euler-Lagrange conditions}
When formulating a fully discrete variational scheme, the time integral in the action functional is split into intervals $[t_n,t_{n+1}]$ (typically of equal lenght) in the manner of
\begin{align}
    %&
    S[\mathbbm{x}(t),\mathbbm{u}(t),\mathbbm{a}(t),\phi(t)]
    %\nonumber
    %\\
    &=\sum_{n=0}^{N-1}\int_{t_n}^{t_{n+1}} L(\mathbbm{x}(t),\dot{\mathbbm{x}}(t),\mathbbm{u}(t),\mathbbm{a}(t),\dot{\mathbbm{a}}(t),\phi(t)) dt.%\nonumber
    %\\
    %=&\sum_{n=0}^{N-1}S_{n,n+1}[\bm{x}(t),u(t),a(t),\phi(t)]
\end{align}
To obtain discrete update maps for the degrees of freedom $(\mathbbm{x}(t), \mathbbm{u}(t),\mathbbm{a}(t),\phi(t))$, one then assumes some discrete representations for the variable paths in the intervals $t\in [t_n,t_{n+1}]$ and computes the time integrals either analytically or with some quadrature rule, depending on how complicated the Lagrangian is. 
%We will take this course of action here but, instead of discretizing everything in one go, we perform the discretization in two steps. First, we discretize only the variables $(a(t),\phi(t))$, leaving $(\bm{x}(t),u(t))$ as continuous trajectories. The purpose of this choice is to (i) illustrate how the charge conservation arises from a discrete Noether symmetry with respect to a discrete electromagnetic gauge transformation and (ii) to demonstrate that the charge conservation law is effectively independent of how the trajectrory $\bm{x}(t)$ is discretized, as long as the discrete version of $\dot{\bm{x}}(t)$ is compatible with the chosen trajectory $\bm{x}$. This choice also leads us to deduct how to obtain at most linearly implicit update for the electric field. In the second phase, we then discretize the trajectories $(\bm{x}(t),u(t))$ The second step that the choice of discretizing the particle trajectory does not affect the derivation of 
%are then approximated typically by a quadrature rule $S_{n,n+1}=\int_{t_n}^{t_{n+1}} L(a(t),\dot{a}(t),\phi(t),\bm{x}(t),\dot{\bm{x}}(t),u(t)) dt$ are approximated by some quadrature rule, often describing the paths of the variables in the interval be. Naturally, we would prefer as explicit update maps as possible and we would like to avoid inverting the Gauss's Law. Let's see if any of this works with the choice

Obviously, there is significant amount of freedom in choosing the discretization. Fortunately some guidelines can be found in the literature that deals with the full-particle Vlasov-Maxwell system. Specifically, the choice of discretization for the interaction part determines whether the discrete action is electromagnetically gauge invariant and if the system has a discrete charge conservation law. In that spirit, we follow \cite{Squire-Qin-Tang-PIC:2012PhPl} and choose our discrete action on the interval $t\in[t_n,t_{n+1}]$ according to 
\begin{widetext}
\begin{align}\label{eq:discrete-action}
    &S_{n,n+1}[\mathbbm{x}_n,\mathbbm{x}_{n+1},\mathbbm{u}_n,\mathbbm{u}_{n+1},\mathbbm{a}_n,\mathbbm{a}_{n+1},\phi_n]\nonumber\\
    &=\Delta t\frac{\varepsilon_0}{2}e^{j_1}_nM^1_{j_1,j_2}e^{j_2}_n-\Delta t\frac{\mu_0^{-1}}{2}(b_n^{k_1}+b_{\text{ext}}^{k_1})M^2_{k_1k_2}(b_n^{k_2}+b_{\text{ext}}^{k_2})
    \nonumber
    \\
    &+\sum_p \Big[q (a_{n+1}^j+a_{\text{ext}}^j)\int_{0}^{1}\bm{W}^1_j(\bm{x}_{p,n}^{n+1}(\tau))\cdot\frac{d\bm{x}_{p,n}^{n+1}(\tau)}{d\tau}d\tau-q\phi_n^iW_i^0(\bm{x}_{p,n})\Delta t\Big]
    \nonumber
    \\
    &+\sum_p\int_{0}^{1}\Big[m u_{p,n}^{n+1}(\tau)\bm{b}_{d,\text{ext}}(\bm{x}_{p,n}^{n+1}(\tau))\cdot\frac{d\bm{x}_{p,n}^{n+1}(\tau)}{d\tau}-E_d(\bm{x}_{p,n}^{n+1}(\tau),u_{p,n}^{n+1}(\tau);\mu_p)\Big]d\tau
    \nonumber
    \\
    &-\sum_p K_d(\bm{x}_{p,n},u_{p,n},\mathbbm{e}_n,\mathbbm{b}_n;\mu_p)\Delta t.
\end{align}
\end{widetext}
In the above expression, the following abbreviations have been introduced
\begin{align}\label{eq:electric-magnetic-field}
    b^k_n&=a^j_n\curl_j^k,\\
    e^j_n&=-(a^j_{n+1}-a^j_{n})/\Delta t-\phi^i_{n}\grad^j_i,\\
    \bm{x}_{p,n}^{n+1}(\tau)&=\bm{x}_{p,n}+\tau(\bm{x}_{p,n+1}-\bm{x}_{p,n}),\\
    u_{p,n}^{n+1}(\tau)&=u_{p,n}+\tau(u_{p,n+1}-u_{p,n}).
\end{align}
In discretizing the guiding-center contribution, the fourth line in \eqref{eq:discrete-action}, several different approaches could have been taken, especially since the fourth line will not affect the charge conservation law. We have chosen the current expression as it will lead to discrete equations for $\mathbbm{x}$ and $\mathbbm{u}$ that are clear analogs of the time-continuous equations of motion. 
%In \eqref{eq:discrete-action}, the discretization of the guiding-center contribution has been chosen to follow the discretization of the interaction term. we have chosen an analogous discretization for the guiding-center contribution to as the analogous  discretization treats the terms $has been done 
%The total discrete action is then given by $S_d=\sum_{n=0}^{N-1}S_{n,n+1}[\mathbbm{a}_n,\mathbbm{a}_{n+1},\phi_n,\mathbbm{x}_n,\mathbbm{x}_{n+1},\mathbbm{u}_n,\mathbbm{u}_{n+1}]$
%\begin{align}
%    S_{n,n+1}&=\Delta t\frac{\varepsilon_0}{2}\left(-\frac{a^{j_1}_{n+1}-a^{j_1}_{n}}{\Delta t}-\phi^{i_1}_{n}\grad^{j_1}_{i_1}\right)M^1_{j_1,j_2}\left(-\frac{a^{j_2}_{n+1}-a^{j_2}_{n}}{\Delta t}-\phi^{i_2}_{n}\grad^{j_2}_{i_2}\right)\nonumber
%    \\
%    &-\Delta t\frac{\mu_0^{-1}}{2}(a_n^{j_1}+\tilde{a}^{j_1})\curl_{j_1}^{k_1}M^2_{k_1k_2}\curl_{j_2}^{k_2}(a_n^{j_2}+\tilde{a}^{j_2})\nonumber
%    \\
%    &+ \sum_p \big[e_p (a_{n+1}^j+\tilde{a}^j)\int_{t_n}^{t_{n+1}}\bm{W}^1_j(\bm{x}_p)\cdot\dot{\bm{x}}_p dt-e_p\phi_n^iW_i^0(\bm{x}_{p,n})\Delta t\big]\nonumber
%    \\
%    &+\sum_p\int_{t_n}^{t_{n+1}}\big[\vartheta_{\alpha}(\bm{z}_p)\dot{z}_p^{\alpha}-(E(\bm{z}_p)+K(\bm{z}_p,e_n,b_n))\big]dt\nonumber
%    \\
%    &\equiv S_{n,n+1}(a_n,a_{n+1},\phi_n,\bm{z}(t))
%\end{align}
%Here we have the abbreviations
%and, for now, we let the trajectory $\bm{z}(t)$ to be arbitrary. 

To derive the discrete Euler-Lagrange conditions, one perturbs the variables, assuming the perturbations to vanish at the end points in time, and looks for a stationary point of the discrete action. With respect to the perturbations $\mathbbm{a}_n\rightarrow \mathbbm{a}_n+\epsilon\delta \mathbbm{a}_n$, this leads to the equation
%\begin{align}
%    \partial_{\epsilon}|_{\epsilon=0}S_d&=\sum_{n=0}^{N-1}\left(\partial_{\epsilon}|_{\epsilon=0}S_{n,n+1}[a_n+\epsilon\delta a_n,a_{n+1},\phi_n)+\partial_{\epsilon}|_{\epsilon=0}S_{n,n+1}(a_n,a_{n+1}+\epsilon\delta a_{n+1},\phi_n)\right)\nonumber
%    \\
%    &=\sum_{n=1}^{N-1}\left(\partial_{\epsilon}|_{\epsilon=0}S_{n,n+1}(a_n+\epsilon\delta a_n,a_{n+1},\phi_n)+\partial_{\epsilon}|_{\epsilon=0}S_{n-1,n}(a_{n-1},a_{n}+\epsilon\delta a_{n},\phi_{n-1})\right)
%\end{align}
%Thus, for each pair $(n,n+1)$, we obtain an equation
\begin{align}
    %&
    \partial_{\epsilon}|_{\epsilon=0}S_{n,n+1}[\mathbbm{a}_n+\epsilon\delta \mathbbm{a}_n]%,\mathbbm{a}_{n+1}]%,\phi_n,\bm{x}_n,\bm{x}_{n+1},u_n,u_{n+1})
    %\nonumber
    %\\
    &+\partial_{\epsilon}|_{\epsilon=0}S_{n-1,n}[%\mathbbm{a}_{n-1},
    \mathbbm{a}_{n}+\epsilon\delta \mathbbm{a}_{n}]=0,%,\phi_{n-1},\bm{x}_{n-1},\bm{x}_{n},u_{n-1},u_{n})=0
\end{align}
and, when written explicitly, provides the discrete analog of the Ampr\`ere-Maxwell equation
\begin{align}\label{eq:discrete-ampere}
    &\varepsilon_0M^1_{j,j_2}\frac{e_n^{j_2}-e^{j_2}_{n-1}}{\Delta t}+J^{n-1,n}_j
    %\nonumber
    %\\
    %&
    +\frac{\mathcal{P}^n_j(\mathbbm{e}_n,\mathbbm{b}_n)-\mathcal{P}^{n-1}_j(\mathbbm{e}_{n-1},\mathbbm{b}_{n-1})}{\Delta t}%\frac{\partial\mathcal{K}_n(\mathbbm{e}_n,\mathbbm{b}_n)}{\partial e^j_n}+\frac{1}{\Delta t}\frac{\partial \mathcal{K}_{n-1}(\mathbbm{e}_{n-1},\mathbbm{b}_{n-1})}{\partial e^j_{n-1}}
    %\nonumber
    %\\
    %&+\curl_{j}^{k}\mathcal{M}^n_{k}(\mathbbm{e},\mathbbm{b})
    \nonumber
    \\
    &=\mu_0^{-1}\curl_{j}^{k} M^2_{k,k_2}(b_n^{k_2}+b_{\text{ext}}^{k_2})-\curl_{j}^{k}\mathcal{M}^n_{k}(\mathbbm{e},\mathbbm{b}),
\end{align}
where the discrete analog of the free current is 
\begin{align}
    J_j^{n,n+1}&=\sum_p q \int_{0}^{1}\bm{W}^1_j(\bm{x}_{p,n}^{n+1}(\tau))\cdot\frac{d\bm{x}_{p,n}^{n+1}(\tau)}{d\tau}\frac{d\tau}{\Delta t},
\end{align}
and the discrete analogs of polarization and magnetization are defined as 
\begin{align}
\mathcal{P}^n_j(\mathbbm{e},\mathbbm{b})&=\sum_p\Big(2K_{j,j_2}^{\mathbbm{ee}}(\bm{x}_{p,n})e^{j_2}+K^{\mathbbm{eb}}_{j,k_2}(\bm{x}_{p,n},u_{p,n})b^{k_2}\Big),\\
\mathcal{M}^n_{k}(\mathbbm{e},\mathbbm{b})&=\sum_p\Big(K_{j_2,k}^{\mathbbm{eb}}(\bm{x}_{p,n},u_{p,n})e^{j_2}-K_k^{\mathbbm{b}}(\bm{x}_{p,n},\mu_p)-2K^{\mathbbm{bb}}_{k,k_2}(\bm{x}_{p,n},u_{p,n},\mu_p)b^{k_2}\Big).
\end{align}
The discrete Amp\`ere-Maxwell equation effectively contains the discrete versions of the polarization and magnetization currents in a manner analogous to the fully continuous system, and is linear in the degrees of freedom $\mathbbm{e}_n$. 

With respect to perturbations $\phi_n\rightarrow\phi_n+\epsilon\delta\phi_n$, the variation of the action leads to
\begin{align}
    \partial_{\epsilon}|_{\epsilon=0}S_{n,n+1}[\phi_n+\epsilon\delta \phi_n]=0,
\end{align}
which, when written explicitly, corresponds to the discrete Gauss' law
\begin{align}\label{eq:discrete-gauss}
    \varrho_i^n&=-\grad_i^j\left(\mathcal{P}^n_j(\mathbbm{e}_n,\mathbbm{b}_n)+\varepsilon_0M^1_{j,j_2}e^{j_2}_n\right),
\end{align}
where the discrete free charge is defined according to
\begin{align}
    \varrho_i^n&=\sum_pq W_i^0(\bm{x}_{p,n}).
\end{align}
Also here it is evident that the discrete Gauss' law contains the analog of polarization density in a manner analogous to the continuous case.

With respect to perturbations in drift-centers' spatial positions, $\mathbbm{x}_n\rightarrow \mathbbm{x}_n+\epsilon\delta\mathbbm{x}_n$, variation of the action provides
\begin{align}
    %&
    \partial_{\epsilon}|_{\epsilon=0}S_{n,n+1}[\mathbbm{x}_{n}+\epsilon\delta\mathbbm{x}_n]
    %\nonumber\\
    &+\partial_{\epsilon}|_{\epsilon=0}S_{n-1,n}[\mathbbm{x}_{n}+\epsilon\delta\mathbbm{x}_n]=0.
\end{align}
Written explicitly, this corresponds to the equation
\begin{align}\label{eq:discrete-el-x}
    &q\frac{\bm{x}_{p,n+1}-\bm{x}_{p,n}}{\Delta t}\times \int_0^1(1-\tau)\bm{B}^{\star}_{d}(\bm{x}_{p,n}^{n+1}(\tau),u_{p,n}^{n+1}(\tau),b_{n+1})d\tau
    \nonumber
    \\
    &q\frac{\bm{x}_{p,n}-\bm{x}_{p,n-1}}{\Delta t}\times \int_0^1\tau\bm{B}^{\star}_{d}(\bm{x}_{p,n-1}^{n}(\tau),u_{p,n-1}^{n}(\tau),b_n)d\tau
    \nonumber
    \\
    %&q\frac{\bm{x}_{p,n+1}-\bm{x}_{p,n}}{\Delta t}\times (b_{n+1}^k+b_{\text{ext}}^k)\int_0^1(1-\tau)\bm{W}^2_k(\bm{x}_{p,n}^{n+1}(\tau))d\tau
    %\nonumber
    %\\
    %&+q\frac{\bm{x}_{p,n}-\bm{x}_{p,n-1}}{\Delta t}\times(b_{n}^k+b_{\text{ext}}^k)\int_0^1\tau\bm{W}^2_k(\bm{x}_{p,n-1}^{n}(\tau)) d\tau
    %\nonumber
    %\\
    %&+m\frac{\bm{x}_{p,n+1}-\bm{x}_{p,n}}{\Delta t}\times\int_{0}^{1}(1-\tau)u_{p,n}^{n+1}(\tau)\nabla\times\bm{b}_{d,\text{ext}}(\bm{x}^{n+1}_{p,n}(\tau))d\tau
    %\nonumber
    %\\
    %&+m\frac{\bm{x}_{p,n}-\bm{x}_{p,n-1}}{\Delta t}\times\int_{0}^{1}\tau u_{p,n-1}^{n}(\tau)\nabla\times\bm{b}_{d,\text{ext}}(\bm{x}^{n}_{p,n-1}(\tau))d\tau
    %\nonumber
    %\\
    &-m\frac{u_{p,n+1}-u_{p,n}}{\Delta t}\int_{0}^{1}(1-\tau)\bm{b}_{d,\text{ext}}(\bm{x}^{n+1}_{p,n}(\tau))d\tau
    \nonumber
    \\
    &-m\frac{u_{p,n}-u_{p,n-1}}{\Delta t}\int_{0}^{1}\tau\bm{b}_{d,\text{ext}}(\bm{x}^{n}_{p,n-1}(\tau),)d\tau
    \nonumber
    \\
    &-\mu_p\int_{0}^{1}\big[(1-\tau)\nabla |\bm{B}_{d,\text{ext}}|(\bm{x}^{n+1}_{p,n}(\tau))+\tau \nabla |\bm{B}_{d,\text{ext}}|(\bm{x}^{n}_{p,n-1}(\tau))\big]d\tau
    \nonumber
    \\
    &+q e_n^j\bm{W}^1_j(\bm{x}_{p,n})-\nabla K_d(\bm{x}_{p,n},u_{p,n},e_n,b_n;\mu_p)=0.
\end{align}
Here the discrete version of the "B-star" field reads
\begin{align}
    \bm{B}^{\star}_{d}&=(b^k+b_{\text{ext}}^k)\bm{W}_k^2+(m/q)u\nabla\times\bm{b}_{d,\text{ext}}.
\end{align}
Finally, the perturbations $\mathbbm{u}_n\rightarrow \mathbbm{u}_n+\epsilon\delta \mathbbm{u}_n$ provide
\begin{align}
    %&
    \partial_{\epsilon}|_{\epsilon=0}S_{n,n+1}[\mathbbm{u}_n+\epsilon\delta \mathbbm{u}_n]
    %\nonumber\\
    %&
    +\partial_{\epsilon}|_{\epsilon=0}S_{n-1,n}[\mathbbm{u}_{n}+\epsilon\delta \mathbbm{u}_n]=0,
\end{align}
which leads to the discrete Euler-Lagrange condition for the drift-center parallel velocity
\begin{align}\label{eq:discrete-el-u}
    &m\frac{\bm{x}_{p,n+1}-\bm{x}_{p,n}}{\Delta t}\cdot\int_{0}^{1}(1-\tau)\bm{b}_{d,\text{ext}}(\bm{x}^{n+1}_{p,n}(\tau))d\tau
    %\nonumber
    %\\
    %&
    +m\frac{\bm{x}_{p,n}-\bm{x}_{p,n-1}}{\Delta t}\cdot\int_{0}^{1}\tau\bm{b}_{d,\text{ext}}(\bm{x}^{n}_{p,n-1}(\tau))d\tau
    \nonumber
    \\
    &=m\int_{0}^{1}\big[(1-\tau)u_{p,n}^{n+1}(\tau)+\tau u_{p,n-1}^{n}(\tau)\big]d\tau
    %\nonumber
    %\\
    %&
    +\partial_uK_d(\bm{x}_{p,n},u_p,e_n,b_n;\mu_p).
\end{align}
%Here we have introduced the abbreviations
%\begin{align}
%    \bm{B}^{\star}_{d}(\bm{x},u,b)&=(b^k+b_{\text{ext}}^k)\bm{W}_k^2(\bm{x})+(m/q)u\nabla\times\bm{b}_{d,\text{ext}}(\bm{x})\\
%    \mathcal{K}_n(e,b)&=\sum_pK_d(\bm{x}_{p,n},u_{p,n},e,b;\mu_p)\\
%    J_j^{n,n+1}&=\sum_p q \int_{0}^{1}\bm{W}^1_j(\bm{x}_{p,n}^{n+1}(\tau))\cdot\frac{d\bm{x}_{p,n}^{n+1}(\tau)}{d\tau}\frac{d\tau}{\Delta t}\\
%    \varrho_i^n&=\sum_pq W_i^0(\bm{x}_{p,n}).
%\end{align}
%Proceeding similarly, the variation of the action with respect to perturbations $\{\phi_n\}_{n=0}^{N}\rightarrow \{\phi_n+\epsilon\delta\phi_n\}_{n=0}^{N}$ leads to
%\begin{align}
%    \partial_{\epsilon}|_{\epsilon=0}S&=\sum_{n=0}^{N-1}\partial_{\epsilon}|_{\epsilon=0}S_{n,n+1}(a_n,a_{n+1},\phi_n+\epsilon\delta \phi_n)=0
%\end{align}
It might be somewhat difficult to interpret what the discrete Euler-Lagrange conditions \eqref{eq:discrete-el-x} and \eqref{eq:discrete-el-u} for $(\mathbbm{x},\mathbbm{u})$ actually represent. Their meaning becomes transparent at the limit $\Delta t\rightarrow 0$ when $\mathbbm{u}_{n+1}\rightarrow \mathbbm{u}_n$, $\mathbbm{u}_{n-1}\rightarrow \mathbbm{u}_n$, $\mathbbm{x}_{n+1}\rightarrow \mathbbm{x}_n$, and $\mathbbm{x}_{n-1}\rightarrow \mathbbm{x}_n$. At this limit, one finds that \eqref{eq:discrete-el-x} reduces to
\begin{align}
    &q \bm{B}_d^{\star}(u_{p,n},\bm{x}_{p,n})\times\dot{\bm{x}}_{p,n}+m\dot{u}_{p,n}\bm{b}_{d,\text{ext}}(\bm{x}_{p,n})\nonumber
    \\
    &=q e_n^j\bm{W}^1_j(\bm{x}_{p,n})-\mu_p\nabla|\bm{B}_{d,\text{ext}}|(\bm{x}_{p,n})-\nabla K_d(\bm{x}_{p,n},u_{p,n},\mathbbm{e}_n,\mathbbm{b}_n;\mu_p)
\end{align}
and that \eqref{eq:discrete-el-u} reduces to
\begin{align}
    m\dot{\bm{x}}_{p,n}\cdot\bm{b}_{d,\text{ext}}(\bm{x}_{p,n})=mu_{p,n}+\partial_uK_d(\bm{x}_{p,n},u_{p,n},\mathbbm{e}_n,\mathbbm{b}_n;\mu_p).
\end{align}
These are exactly the conditions from which the time-continuous equations of motion would be recovered for $(\dot{\bm{x}}_{p,n},\dot{u}_{p,n})$, after taking a cross-product with $\bm{b}_{d,\text{ext}}$ and a dot-product with respect to $\bm{B}_{d}^{\star}$. However, starting from the time-continuous equations of motion, it might be difficult to guess such forms for the discrete versions without help from a variational principle. 

The equations \eqref{eq:discrete-ampere}, \eqref{eq:discrete-gauss}, \eqref{eq:discrete-el-x}, and \eqref{eq:discrete-el-u} are to be completed by the discrete Faraday equation that is a direct consequence of the definitions for $\mathbbm{e}_n,\mathbbm{b}_n$, namely
\begin{align}\label{eq:discrete-faraday}
    \frac{b^k_{n}-b_{n-1}^k}{\Delta t}=-\curl_j^ke_{n-1}^j.
\end{align}
Together the discrete equations provide means of advancing the degrees of freedom $\mathbbm{x}_n$, $\mathbbm{u}_n$, $\mathbbm{e}_n$, and $\mathbbm{b}_n$ in time according to the following strategy
\begin{enumerate}
    \setcounter{enumi}{-1}
    \item Initialize with Gauss law \eqref{eq:discrete-gauss} $(\mathbbm{b}_0,\mathbbm{x}_0,\mathbbm{u}_0)\rightarrow \mathbbm{e}_0$ and approximate $(\mathbbm{x}_{-1},\mathbbm{u}_{-1})$
    \item Advance Faraday equation \eqref{eq:discrete-faraday}: $(\mathbbm{e}_n,\mathbbm{b}_n)\rightarrow \mathbbm{b}_{n+1}$
    \item Push markers with \eqref{eq:discrete-el-x} and \eqref{eq:discrete-el-u}: $(\mathbbm{e}_n,\mathbbm{b}_n,\mathbbm{b}_{n+1},\mathbbm{x}_{n-1},\mathbbm{x}_{n},\mathbbm{u}_{n-1},\mathbbm{u}_{n})\rightarrow (\mathbbm{x}_{n+1},\mathbbm{u}_{n+1})$
    \item Invert Amp\`ere-Maxwell equation \eqref{eq:discrete-ampere}: $(\mathbbm{e}_n,\mathbbm{b}_n,\mathbbm{b}_{n+1},\mathbbm{x}_n,\mathbbm{x}_{n+1},\mathbbm{u}_n,\mathbbm{u}_{n+1})\rightarrow \mathbbm{e}_{n+1}$
    \item Repeat steps 1-3 for $n=0,...,N$.
\end{enumerate}
In the above algorithm, the Gauss' law is to be inverted only once. This is enough as it will be satisfied at later times automatically, as we will demonstrate next.

\section{Discrete charge conservation and Gauss' law}
As a final step before summarizing our results, we analyze the electromagnetic gauge invariance of the fully discrete action. Effectively, we let 
\begin{align}
    &a^j_n\rightarrow a^j_n+\chi^i_n\grad^j_i, \\ &\phi^i_n\rightarrow\phi^i_n-\frac{\chi^i_{n+1}-\chi^i_{n}}{\Delta t}.
\end{align}
By their definition, $\mathbbm{e}_n$ and $\mathbbm{b}_n$ are invariant under these changes, and we observe that the discrete action \eqref{eq:discrete-action} changes according to 
\begin{align}
S_{n,n+1}\rightarrow &S_{n,n+1}+\sum_p q\Big[ \chi_{n+1}^i\grad_i^j\int_{0}^{1}\bm{W}^1_j(\bm{x}_{p,n}^{n+1}(\tau))\cdot\frac{d\bm{x}_{p,n}^{n+1}(\tau)}{d\tau}d\tau+(\chi_{n+1}^i-\chi_n^i)W_i^0(\bm{x}_{p,n})\Big],\nonumber
\\
=&S_{n,n+1}+\sum_pq\big[ \chi_{n+1}^iW_i^0(\bm{x}_{p,n+1})-\chi^i_{n}W_i^0(\bm{x}_{p,n})\big].
\end{align}
When summed over different $n$, the extra terms only produce pure temporal boundary terms
\begin{align}
    \sum_{n=0}^{N-1}(\chi_{n+1}^iW_i^0(\bm{x}_{p,n+1})-\chi^i_{n}W_i^0(\bm{x}_{p,n}))=\chi_{N}^iW_i^0(\bm{x}_{p,N})-\chi^i_{0}W_i^0(\bm{x}_{p,0}).
\end{align}
When the action is varied, the variations of the first and last points $\bm{x}_{p,0}$ and $\bm{x}_{p,N}$ are held fixed and hence this change in the gauge does not alter the resulting Euler-Lagrange conditions. In this sense, the discrete action is invariant with respect to the discrete gauge transformation, up to the temporal boundary terms.

This discrete invariance of the action is an analog of a continuous Noether symmetry of the infinite-dimensional drift-kinetic action functional under the change of electromagnetic gauge. And exactly similarly as in the continuous case, the discrete gauge invariance provides the discrete charge-conservation law. This is seen once the explicit form of the invariance condition, namely
\begin{align}
    &\sum_{n=0}^{N-1}S_{n,n+1}(a^j_n+\chi^i_n\grad_i^j,a^j_{n+1}+\chi^i_{n+1}\grad_i^j,\phi^i_n-(\chi^i_{n+1}-\chi^i_n)/\Delta t,\mathbbm{x}_n,\mathbbm{x}_{n+1},\mathbbm{u}_n,\mathbbm{u}_{n+1})\nonumber
    \\
    &=\sum_{n=0}^{N-1}S_{n,n+1}(\mathbbm{a}_n,\mathbbm{a}_{n+1},\phi_n,\mathbbm{x}_n,\mathbbm{x}_{n+1},\mathbbm{u}_n,\mathbbm{u}_{n+1})+\sum_p e_p\big[\chi_{N}^iW_i^0(\bm{x}_{p,N})-\chi^i_{0}W_i^0(\bm{x}_{p,0})\big],
\end{align}
is differentiated with respect to $\chi_{n}$ at any $n$ such that $n\neq 0$ and $n\neq N$. The right side vanishes identically as it is independent of $\chi_{n}$, and we find
\begin{align}
    \grad_i^jJ_{j}^{n-1,n}-\frac{\varrho^{n}_i-\varrho_i^{n-1}}{\Delta t}=0,% \qquad \forall\quad  \tilde{n}=1,...,N-1
\end{align}
To obtain this result, only the matrix identity $\grad_i^j\curl_j^k=0$ has been used, everything else exactly cancels out. 

To see the significance of this equation, we assume the Gauss' law \eqref{eq:discrete-gauss} to hold for $n-1$. The charge conservation and \eqref{eq:discrete-ampere} then imply
\begin{align}
    \varrho^{n}_i=&\varrho_i^{n-1}+\Delta t\,\grad_i^jJ_{j}^{n-1,n}=-\grad_i^j\left(\mathcal{P}^n_j(\mathbbm{e}_n,\mathbbm{b}_n)+\varepsilon_0M^1_{j,j_2}e^{j_2}_n\right),
\end{align}
meaning that the Gauss' law is automatically satisfied, if it is satisfied initially. This property is analogous to fully continuous system, where the Gauss' law serves as an initial condition for the infinite-dimensional system.

\section{Summary and discussion}
This paper was devoted to investigating the possibility of a variational algorithm for the drift-kinetic Vlasov-Maxwell system. As it was demonstrated, such a discrete scheme was indeed found and, furthermore, guarantees a discrete charge-conservation law as a consequence of the discrete electromagnetic gauge invariance of the action. The key to the presented results was the recent discovery of a gauge-free electromagnetic gyrokinetic theory which allows one to express the drift-kinetic perturbation terms in the action in terms of the perturbed electric and magnetic field instead of the perturbed electromagnetic potentials. Consequently, also the discrete equations involve only the electromagnetic fields. Perhaps the most important effect of this fact is that it opens up the possibility to construct a structure-preserving variational scheme that couples fully kinetic ions and drift-kinetic electrons to study electromagnetic turbulence and the associated transport in steep background gradients where the gyrokinetic assumptions for ions might not necessarily be valid.

Finally, we stress that our choice for the discretization is not unique. The only guiding principle was to retain the fully discrete action gauge invariant. For example, the choice for discretizing the guiding-center part of the action was based purely on the aesthetically appealing looks of the resulting discrete equations for advancing the individual drift-center coordinates. Future studies into the topic should focus on discretizations that contain only polynomial dependencies on the drift-center degrees of freedom for efficient numerical integration of the line integrals and, especially, on the possible stability issues in degenerate variational phase-space discretizations. In the end, this paper presents only the first attempt at constructing a structure-preserving integrator for the drift-kinetic plasma model. Hopefully more will come, and the superior long-time stability properties of the new algorithms find their way to production-level codes within the fusion-research community. At least one such code is to be launched with the next years, to reboot \href{http://elmfire.eu}{ELMFIRE} full-$f$ gyrokinetic programme currently developed and maintained at Aalto University.

\begin{acknowledgments}
The author is grateful to Joshua W. Burby and Alain J. Brizard for the numerous discussions over the years regarding action principles and the gyrokinetic theory, and for the encouragement from the ELMFIRE group. Financial support for the research was provided by the Academy of Finland grant no. 315278. Any subjective views or opinions expressed herein do not necessarily represent the views of the Academy of Finland or Aalto University.
\end{acknowledgments}

\bibliography{references}  
\end{document}